\documentclass[12pt]{article}
\usepackage{amsmath,amssymb,amsthm,amsxtra,overpic,bbm,bm,epsfig,ulem}
\usepackage{color}
\usepackage{cite}

\usepackage{hyperref}
\usepackage{url}
\usepackage{comment}

\usepackage{tikz-feynman}
\usepackage{tikz}
\usetikzlibrary{decorations.pathmorphing}
\usetikzlibrary{decorations.pathmorphing} 
\tikzset{graviton/.style={decorate, decoration={snake, amplitude=.4mm, segment length=1.5mm, pre length=.5mm, post length=.5mm}, double}}

\def\thefootnote{\fnsymbol{footnote}}

\begin{document}

\vspace{0.2cm}

\begin{center}
{\large\bf  Graviton energy spectra arising from the  KSVZ axion model}
\end{center}

\vspace{0.2cm}

\begin{center}
{\bf Yonghua Wang }$^{1,2}$
\footnote{E-mail: yonghuawang@mail.bnu.edu.cn}
{\bf Lin-Yun He }$^{1,2}$
\footnote{E-mail: linyunhe@mail.bnu.edu.cn}
\footnote{HLY and WYH contribute equally to the project.}
{\bf Wei Chao }$^{1,2}$
\footnote{E-mail: chaowei@bnu.edu.cn}
{\bf Yu Gao}$^{3}$
\footnote{E-mail: gaoyu@ihep.ac.cn}
\\
{\small $^{1}$Key Laboratory of Multi-scale Spin Physics, Ministry of Education, Beijing Normal University, Beijing 100875, China} \\
{\small $^{2}$Center of Advanced Quantum Studies, School of Physics and Astronomy, Beijing Normal University, Beijing, 100875, China}  \\
{\small $^3$Key Laboratory of Particle Astrophysics, Institute of High Energy Physics, Chinese Academy of Sciences, Beijing 100049, China}
\end{center}

\vspace{1cm}

\begin{abstract}
Axion, the goldstone boson arising from the spontaneous breaking of a global $U(1)$ Peccei-Quinn symmetry,  provides a dynamical solution to the strong CP problem and is an excellent dark matter candidate. Various experiments are designed to search for the axion, however no confirmative signal has been observed. On the other hand, there are also hypothetical heavy particles in axion models, such as the heavy scalar $s$, which is  the CP-even component of the complex scalar that carries $U(1)_{PQ}$ charge,  and the vector-like heavy quark (VLQ) in the Kim-Shifman-Vainshtein-Zakharov~(KSVZ) axion model. Studying signals induced by them are helpful for axion searches. In this paper, we calculate the graviton bremsstrahlung energy spectrum arising from the decay of the heavy scalar or  VLQ in the KSVZ model. The result shows that these heavy particles can emit ultrahigh-frequency gravitational waves (GWs), with the peak frequency depending on the model's parameter inputs.  In addition, the graviton spectrum is distinguished from the thermal GW background at high frequencies if there is an early matter-dominated era induced by these heavy particles. Future measurements of  ultrahigh-frequency GWs may provide  indirect evidence for the KSVZ axion.
\end{abstract}

\newpage

\def\thefootnote{\arabic{footnote}}
\setcounter{footnote}{0}

\section{Introduction}
Theoretically, there are at least two sources of CP violation that contribute to the electric dipole moment (EDM) of nucleons. One is the topological term arising from the ``$\theta$-vacua" of  Quantum Chromodynamics (QCD) and the other one is the phase in the quark matrices $M_u$ and $M_d$. Both terms can be ${\cal O}(1)$ numbers and the combination of them gives an effective CP-violating phase $\bar \theta$. Experimental searches for the EDM of nucleons have put strong constraints on the $\bar \theta$, $\bar \theta = \theta +\arg {\rm det} M_u M_d  \leq 10^{-10} $, which means there is a strong cancellation between these two terms.  This is the so-called strong CP problem, which aims to understanding this fine tune problem.

A popular solution to the strong CP problem is the Peccei-Quinn mechanism~\cite{Peccei:1977hh,Peccei:1977ur,Weinberg:1977ma,Wilczek:1977pj}, in which a global $U(1)_{PQ}$ symmetry is spontaneously broken, giving rise to an axion field $\phi$ and a classical shift symmetry under $\phi \to \phi + constant$.   If the shift symmetry is only violated by the quantum effect as $({\cal C}\phi/f_a) G\tilde G$, then the potential of $\phi$ induced by the QCD non-perturbative effect dynamically set $\bar \theta =0$. Notably, the axion provides a  promising cold dark matter candidate, which makes the axion physics a prominent research direction.  There are three typical axion models: the Peccei–Quinn–Weinberg–Wilczek(PQWW)~\cite{Peccei:1977hh,Wilczek:1977pj,Vafa:1984xg}, the Kim–Shifman–Vainshtein–Zakharov (KSVZ)~\cite{Kim:1979if,Shifman:1979if} and the Dine–Fischler–Srednicki–Zhitnitsky (DFSZ)~\cite{Dine:1981rt,Zhitnitsky:1980tq}. The PQWW axion coupling is too large and is already excluded by  beam dump experiments~\cite{Kim:1986ax}. The KSVZ axion and DFSZ axion are  viable and well-motivated frameworks, but differing in symmetry-breaking mechanisms, couplings to the SM as well as current observational constraints. Axion couplings are constrained by cosmology, astrophysics and particle physics observations.  In addition, axions can be tested in various precision measurement experiments. Non-observation of any signal puts more and more severe upper limits on axion couplings.

Since the observation of GW in the ground-based interferometer~\cite{LIGOScientific:2016aoc}, GW has been taken as a new probe of physics in the early universe. 
Currently, there are many kinds of GW detection facilities or proposals that may detect various frequency bands of GW, typically Pulsar Timing Array (PTA) experiments are capable of detecting GWs in the nanohertz regime, space  based interferometer experiments may detect stochastic GWs in the (0.01 $mHz$, 10 $Hz$) regime, ground based interferometers can detect GWs in the (1 $Hz$, 10 $kHz$) regime and cavity experiments may detect higher frequency GWs.  
In the early universe, GW can be emitted in various processes. Especially, the thermal GW background~\cite{Ghiglieri:2015nfa,Ringwald:2022xif,Ghiglieri:2022rfp,Ghiglieri:2024ghm,Ringwald:2020ist} produced by microscopic collisions of the SM particles in the thermal bath, reflects the energy distribution of the SM plasma. 
High frequency GWs arising from inflation~\cite{barman2023gravitational,Xu:2025wjq,huang2019stochastic,Barman:2023rpg,Bernal:2023wus,Xu:2024fjl,Xu:2024xmw,Bernal:2024jim,Bernal:2025lxp,Tokareva:2023mrt,Kanemura:2023pnv,Montefalcone:2025gxx,Ema:2020ggo,Bernal:2023wus,Klose:2022knn,Klose:2022rxh}, GUT~\cite{An:2022cce,Hu:2025xdt,Chao:2017ilw}, freeze-in dark matter~\cite{Chao:2017vrq,Chao:2023lox,Wang:2025lmf,Konar:2025iuk},  seesaw mechanisms~\cite{Wang:2025mtq}, and cosmic phase transitions~\cite{Caprini:2009yp,Chao:2017vrq,Hindmarsh:2017gnf,Athron:2023xlk,Chao:2021xqv} had been systematically  investigated. 
In this paper, we study high frequency gravitational wave (GW) signal induced by the axion.  We focus on the KSVZ axion model and calculate the graviton bremsstrahlung  during the the decay of the heavy CP-even scalar and heavy vector-like quarks (VLQ) in the early universe. 
We first simulate the evolution of heavy scalar and VLQ, then derive the  GW spectrum from their decays.   
The GW spectrum shows a peak frequency at $10^{10}$ Hz or higher with specific value depending on the model parameters. Future detection of high frequency GWs may provide an indirect  signal of axion model.

The remaining of the paper is organized as follows: In section II we give a brief introduction to the KSVZ axion model; section III is devoted to the study of cosmological evolutions of the heavy scalar and the VLQ; We derive the gravitation spectrum in section IV; The last part is concluding remarks.

\section{The KSVZ model}

The KSVZ model extends the Standard Model (SM) with a vector-like fermion $Q_L$ and $Q_R$, which carry the same  $SU(3)_C$ charge but transform differently under the global PQ symmetry $U(1)_{\rm PQ}$~\cite{PhysRevLett.43.103,SHIFMAN1980493}, and a complex scalar field $\Phi$, which is  singlet under the SM gauge group but carries a non-zero charge under the $U(1)_{\rm PQ}$. The relevant Lagrangian is given by~\cite{DiLuzio:2017pfr}
\begin{equation}
    \mathcal{L} _a = \mathcal{L}_{SM} + \mathcal{L}_{PQ} - V_{\Phi} + \mathcal{L}_{\Phi H} + \mathcal{L}_{Qq},
\end{equation}
where $\mathcal{L}_{SM}$ is the SM Lagrangian, and
\begin{equation}
    \mathcal{L}_{PQ} = |\partial_\mu \Phi|^2 + i \bar{Q} \gamma^\mu D_\mu Q - (y_Q \bar{Q}_L \Phi Q_R + h.c.).
\end{equation}
The potential that triggers the spontaneous breaking of the $U(1)_{\rm PQ}$ symmetry is given by
\begin{equation}
	V_{\Phi} = -\mu_\Phi^2 |\Phi|^2 + \lambda_\Phi |\Phi|^4.
\end{equation}
The term $\mathcal{L}_{\Phi H}$ accounts for the interactions between the complex scalar field $\Phi$ and the SM Higgs doublet $H$ and is given as

\begin{equation}
\mathcal{L}_{\Phi H} = - \lambda_{\Phi H} |\Phi|^2 |H|^2,
\end{equation}
where the coupling $\lambda_{\Phi H}$ is severely constrained by the electroweak symmetry breaking scale.
$\mathcal{L}_{Qq}$ represents the possible interactions between $Q_{L,R}^{} $ and the SM quarks, which lead to the decay of VLQ into SM quarks.
Some interaction operators are phenomenologically preferred, provided that they satisfy the following two criteria~\cite{DiLuzio:2017pfr}:
$ i) $ the lifetime of VLQ is sufficiently short to avoid various cosmological problems, $\tau_Q \lesssim 10^{-2} \text{s} $.
$ii)$ Landau poles (LP) in the SM gauge couplings are avoided below the Planck scale, i.e., $\Lambda_{\rm LP}^{} \gtrsim 10^{18} \text{ GeV}$. 
%
%
Some possible renormalizable operators $\mathcal{O}_{Qq}$ that satisfy the above two criteria are summarized in the Table.~\ref{tab:Qrep}~\cite{DiLuzio:2017pfr}.
\begin{table}[h]
    \centering
    \begin{tabular}{|c|c|c|}
    \hline
        $R_Q$ & $\mathcal{O}_{Qq}$ & $\Lambda_{LP}[GeV]$ \\
        \hline
         $R_1: (3,1,-\frac{1}{3})$ & $\overline{Q}_L d_R$ & $9.3 \cdot 10^{38}(g_1)$ \\
    \hline
    $R_2: (3,1,+\frac{2}{3})$ & $\overline{Q}_L u_R$ & $5.4 \cdot 10^{34}(g_1)$ \\
    \hline
    $R_3: (3,2,+\frac{1}{6})$ & $\overline{Q}_R q_L$ & $6.5 \cdot 10^{39}(g_1)$ \\
    \hline
    $R_4: (3,2,-\frac{5}{6})$ & $\overline{Q}_L d_R H^\dagger$ & $4.3 \cdot 10^{27}(g_1)$ \\
    \hline
    $R_5: (3,2,+\frac{7}{6})$ & $\overline{Q}_L u_R H$ & $5.6 \cdot 10^{22}(g_1)$ \\
    \hline
    $R_6: (3,3,-\frac{1}{3})$ & $\overline{Q}_R q_L H^\dagger$ & $5.1 \cdot 10^{30}(g_2)$ \\
    \hline
    $R_7: (3,3,+\frac{2}{3})$ & $\overline{Q}_R q_L H$ & $6.6 \cdot 10^{27}(g_2)$ \\
    \hline
    \end{tabular}
    \caption{Possible renormalizable operators and representations of the vector-like quark $Q$ under the SM gauge group that satisfy the phenomenological criteria~\cite{DiLuzio:2017pfr}.}
    \label{tab:Qrep}
\end{table}

The complex scalar field $\Phi$ is parameterized as
\begin{equation}
    \Phi(x) = \frac{1}{\sqrt{2}} ( s+ f_a ) e^{i a / f_a }.
\end{equation}
where $f_a$ denotes the PQ symmetry breaking scale, $s$ is the radial excitation of $\Phi$, and the angular excitation, $a$, corresponds to the QCD axion.
Upon the spontaneous breaking of the $U(1)_{\rm PQ}$ symmetry, $Q$ and $s$ attain masses given by $m_Q = y_Q f_a / \sqrt{2}$ and $m_s = \sqrt{2 \lambda_\Phi} f_a$, respectively.
%
%
Interactions of $a$ are given as
\begin{equation}
     - \frac{y_Q} {\sqrt{2}} s \bar{Q} Q - \mu_{\Phi H} s |H|^2,
	 \label{eq:yukawa}
\end{equation}
where $\mu_{\Phi H} \equiv \lambda_{\Phi H} f_a$.
Consequently,  the heavy scalar $s$ possesses two competing decay channels: one into a pair of VLQ and another one into a pair of Higgs doublet.

\section{Thermal history}
In this section, we examine the cosmological evolution of $s$ and $Q$. We trace their dynamics chronologically, following the timeline of the universe from the PQ symmetry breaking through the subsequent decay stages.
We focus on the post-inflationary PQ symmetry breaking scenario~\cite{Ringwald:2015dsf}, where the PQ symmetry is broken after  the reheating, i.e., $T_{\rm rh} > f_a$ with $T_{\rm rh}$ the reheating temperature of the universe.
We assume that $s$ and $Q$ reach thermal equilibrium with the SM plasma after the reheating. This condition is satisfied provided that the interaction rate of the particle in the plasma, $\Gamma$, exceeds the Hubble expansion rate, $H$, at the end of the reheating phase. 
Parametrically, this requirement translates into $\zeta > \sqrt{T_{\text{rh}}/m_{\text{pl}}}$, where $\zeta$ denotes the relevant coupling constant and $m_{\text{pl}}$ is the Planck mass. 
%
%
The general  energy density of  the particle ``$i$" in the plasma is given by \cite{Baumann_2022}
\begin{align}
	\rho_i (T) = \frac{g_i}{2 \pi^2} T^4 \int_0^\infty d \xi \frac{\xi^2 \sqrt{\xi^2 + x^2}}{\exp[\sqrt{\xi^2 + x^2}] \pm 1} ,
\end{align}
where $g_i$ is the internal degrees of freedom, $\xi \equiv p/T$ with $p$  the momentum, $x \equiv m/T$ with $m$ the mass of the particle. The plus and minus signs correspond to  the fermion and the boson separately. Throughout this work, we adopt a notation where a bare symbol represents a physical quantity common to all particle species, while a subscripted symbol denotes the quantity corresponding to a specific species, e.g., $m$ represents mass in general, while $m_s$ and $m_Q$ denote the masses of $s$ and $Q$, respectively.

After the PQ symmetry breaking, interactions  given in Eq.~\eqref{eq:yukawa}  lead to the decay of $s$, and the VLQ can decay into the SM quarks and the Higgs through interactions listed in Table.~\ref{tab:Qrep}. We assume that the decays of particles mainly occur when they are non-relativistic, which implies $T_D < T_m$, where $T_m=m$ is the temperature when the particle becomes non-relativistic, and $T_D$ is the decay temperature that is estimated by~\cite{Hooper:2024avz,Kolb:1990vq}
\begin{equation}
	H(T_D)= \left(\frac{8 \pi G}{3} \frac{g_*^\rho \pi^2}{30} T_D^4 \right)^{1/2} \simeq \frac{2}{3} \Gamma,
	\label{Eq.TD}
\end{equation}
where $\Gamma$ is the decay rate of the related particle, $g_*^\rho$ is the effective number of relativistic degrees of freedom contributing to the energy density. The decay rates of $s$ and $Q$ are given by
\begin{align}
	\Gamma_{sQ} (s \to \bar{Q}+ Q) &= \mathcal{I}_Q \mathcal{C}_Q \frac{y_Q^2 M_s}{16 \pi} \left(1-\frac{4 M_Q ^2}{M_s^2}\right)^{3/2}, \\
	\Gamma_{sH} (s \to H+ H) &= \frac{1}{8 \pi} \frac{\mu_{\Phi H}^2}{M_s}, \\
	\Gamma_Q (Q \to H + q) &= \frac{y_{Qq}^2 M_{Q}}{32 \pi} .
\end{align}
where $y_{Qq}$ is Yukawa coupling related to VLQ decay, $\mathcal{I}_Q$ and $\mathcal{C}_Q$ are the isospin and color degrees of freedom of the vector-like quark, respectively, that depend on the representation of the VLQ under the SM gauge group. If the $s$ mainly decays into VLQs, the condition $T_D < T_m$ leads to the constraints on the couplings $y_Q$ and $\mu_{Qq}$ as
\begin{align}
	y_Q &< 3.6 \times 10^{-3} \left(\frac{m_s}{10^{12} \text{GeV}}\right)^{1/2}, \label{Eq.yQ}\\
	y_{Qq} &< 3.2 \times 10^{-4} \left(\frac{f_a}{10^{12} \text{GeV}}\right)^{1/2} \left(\frac{y_Q}{10^{-3}}\right)^{1/2}.
\end{align}
In the numerical estimation of Eq.~\eqref{Eq.yQ}, we have set $\mathcal{I}_Q = 2$, $\mathcal{C}_Q = 3$ and assumed that $m_s \gg m_Q$. The bound on $y_Q$ requires $m_Q$ to be much lower than the PQ scale; nevertheless, this restriction is alleviated in scenarios where the $s$ decay is dominated by the Di-Higgs channel.

The evolution equation for the energy density of  $s$ after the PQ symmetry breaking is given by
\begin{equation}
	\frac{d \rho_s}{d t} + 3 H \rho_s = - \Gamma_s \rho_s \; ,
\end{equation}
which leads to the following solution,
\begin{equation}
	\rho_s (t) = \rho_s (t_{0}) \left(\frac{a(t_{0})}{a(t)}\right)^3 \text{e}^{-\Gamma_s (t - t_{0})},
\end{equation}
where $t_{0}$ is the start time of the decay, and $a(t)$ is the scale factor of the universe. The produced VLQs from the decay rapidly thermalize with the SM plasma via  the strong interaction. Thus, VLQ keeps in thermal plasma until the temperature of the universe drops below its mass $m_Q$, at which it starts to decay. The evolution equation for the energy density of the vector-like quark when $T< m_Q$ is given by
\begin{equation}
	\frac{d \rho_Q}{d t} + 3 H \rho_Q = - \Gamma_Q \rho_Q.
\end{equation}
The solution to this equation is
\begin{equation}
	\rho_Q (t) = \rho_Q (t_{m_Q}) \left(\frac{a(t_{m_Q})}{a(t)}\right)^3 \text{e}^{-\Gamma_Q (t - t_{m_Q})}.
\end{equation}

\begin{figure}[t]
	\centering
	\includegraphics[width=0.6\textwidth]{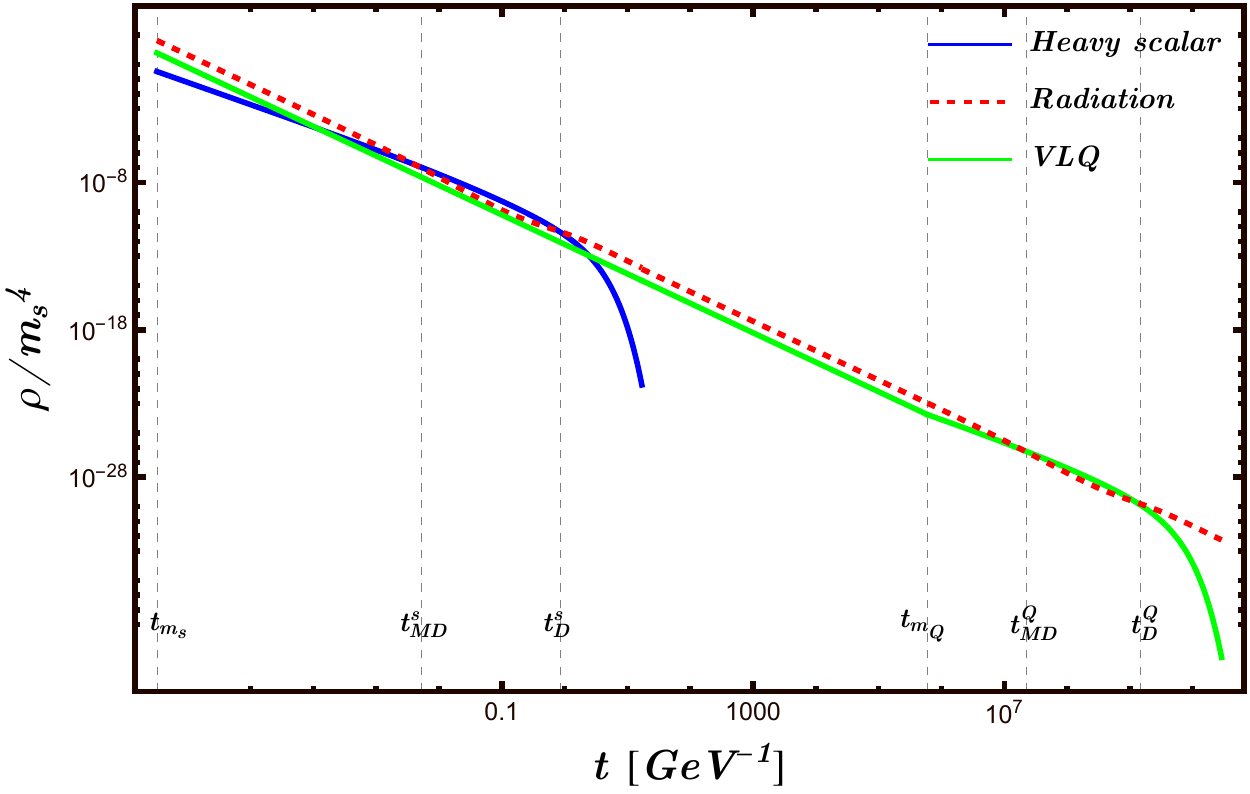}
	\caption{The evolution of the energy densities of the radiation (Red), $s$ (Blue) and $Q$ (Green), in the scenario where $s$ mainly decays to VLQs. We have set $m_s=f_a = 10^{12} \text{ GeV}$, $y_Q = 10^{-6}$, and $y_{Qq} = 10^{-7}$. }
	\label{fig:rho_evolution}
\end{figure}

We depict the evolution of the energy densities of $s$, $Q$ and the SM plasma in Fig.~\ref{fig:rho_evolution}.
It can be seen that the universe undergoes two steps of matter-domination followed by radiation-domination due to heavy particles decay. 
These situations occur when $s$ and $Q$ have a sufficiently long lifetime, which can be achieved by requiring small couplings $y_Q$ and $y_{Qq}$. 
The matter-dominated phase is attained when the energy density of matter equals that of radiation, i.e., $\rho_M(t_{MD}) = \rho_{R}(t_{MD})$, with $t_{MD}$ the time of matter-radiation equality that implies the beginning of the matter-dominated universe and it must happen before the decay of the matter, i.e., $t_{MD} < t_D$. This condition translates to constraints on the couplings $y_Q$ and $y_{Qq}$ as
\begin{align}
	y_Q < & 2.8 \times 10^{-5} \left(\frac{m_s}{10^{12} \text{GeV}}\right)^{1/2}, \\
	y_{Qq} < & 5.3 \times 10^{-6} \left(\frac{f_a}{10^{12} \text{GeV}}\right)^{1/2} \left(\frac{y_Q}{10^{-5}}\right)^{1/2}.
\end{align}
Here we have set $\mathcal{I}_Q = 2$, $\mathcal{C}_Q = 3$ and  $m_s \gg m_Q$. 

The appearance of the matter-dominated phase indicates that the signals of the stochastic gravitational waves generated from the matter decay can be distinguished from the stochastic gravitational wave background produced by the thermal plasma in the high frequency regime~\cite{murayama2025observing}.

\section{Gravitational wave spectrum}
In this section, we will calculate the GW spectra generated from graviton bremsstrahlung in decays of $s$ and $Q$, and compare them with those from the SM thermal plasma \cite{Ringwald:2020ist}. The Feynman rules for graviton-matter interactions can be found in \cite{Wang:2025mtq,Wang:2025lmf,murayama2025observing,barman2023gravitational}.

The GW spectrum today is defined as
\begin{equation}
	h^2 \Omega_{GW} (f) \equiv  h^2 \frac{1}{\rho_{cr,0}} \frac{d \rho_{GW}^0}{d \ln f},
\end{equation}
where $h^2$ is a factor eliminating the uncertainty in the Hubble parameter, $\rho_{cr,0} $ is the critical energy density of the universe today, $\rho_{GW}^0$ is the energy density of GW today, and $f$ is the frequency of GW today. 
%
The evolution of the energy density of GWs is described by the Boltzmann equation
\begin{equation}
	\frac{d \rho_{GW}}{dt} + 4 H \rho_{GW} = \mathcal{C}, \label{Eq.BE}
\end{equation}
where $\mathcal{C}$ is the collision term that describes the creation and annihilation of graviton. The general expression of the collision term for graviton bremsstrahlung process in a 2-body decay is given by\cite{Hooper:2024avz,Kolb:1990vq}
\begin{equation}
	\begin{aligned}
	\mathcal{C} = &g_X \int d\Pi_X f_X d\Pi_i (1 \pm f_i) d\Pi_j (1 \pm f_j) d\Pi_g (1 \pm f_g) (2\pi)^4 \delta^4(P_X - P_i - P_j - P_g) | \mathcal{M}_{X \to ijg} |^2 E_g,
	\end{aligned}
\end{equation}
where $g_X$ is the internal degrees of freedom of the decaying particle $X$, $E_g$ is the energy of the graviton, and the indices $i,j$ denote  two final states other than the graviton. $f_X = (\text{e}^{E_X/T} \pm 1)^{-1}$ is the distribution function with $\pm$ for fermion and boson separately, and $d \Pi = \frac{d^3 p}{(2\pi)^3 2 E} $ is the Lorentz invariant phase space element. The squared amplitude of the decay process is represented by $| \mathcal{M}_{X \to ijg} |^2$. We have neglected the back reaction from graviton annihilation as the graviton population is sufficiently small.

The integral of the distribution function $f_X$ over the phase space is given by
\begin{equation}
	g_X \int d\Pi_X f_X = \frac{1}{2 m_X} n_X \frac{K_1(m_X/T)}{K_2(m_X/T)}, \label{Eq.n}
\end{equation}
where $n_X$ is the number density, $m_X$ is the mass of the particle, and $K_1$ and $K_2$ are the modified Bessel functions of the second kind. We have assumed that particle $X$ keeps in kinetic equilibrium when it decays in the derivation of Eq.~\ref{Eq.n}. The factor of Bessel functions $\frac{K_1(m_a/T)}{K_2(m_a/T)} \sim 1$ when $T \ll m_X$. The integral of the delta function over the 3-body final states phase space gives\cite{Maggiore_book}
\begin{equation}
	\int d\Pi_i \Pi_j d\Pi_g (2\pi)^4 \delta^4(P_a - P_i - P_j - P_g) = \frac{1}{32 \pi^3} \int dE_j dE_g.
\end{equation}
To obtain the GW spectrum today, we differentiate the Boltzmann equation with respect to the graviton energy $E_g$ and integrate over time from $t_m$, the time when the decaying particles become non-relativistic, to $t_D$, the time corresponding to the decay temperature $T_D$ at which the mother particle has basically decayed out.

The final ingredient for the GW-spectrum is the squared amplitude $|\mathcal{M}_{X \to ijg}|^2$. In the following sections, we will focus on the specific $s$ and $Q$ decay processes, calculate their corresponding decay amplitudes and obtain final GW spectra. These two decay stages can be treated independently. If $s$ decays primarily into VLQs, the small Yukawa coupling $y_Q$—required for the existence of a matter-dominated phase—ensures that the $s$ decay process occurs significantly earlier than that of the VLQ decay. Alternatively, if $s$ decays predominantly into Higgs doublets, the two processes are entirely independent. Therefore, their GW signatures can be treated separately, and the total GW spectrum is the sum of the two contributions.

\subsection{GW-spectrum arising from $s$ decay}
In this section, we calculate the GW spectrum from graviton bremsstrahlung processes in the decay of $s$. There are two decay channels, $s\to Q +\bar Q +g$ and $s\to H+ H+g$ where $g$ is the graviton.

\subsubsection{VLQs decay channel}
The corresponding Feynman diagrams for $s\to Q +\bar Q +g$ process are shown in the first row of the Fig.~\ref{fig:feynman}.
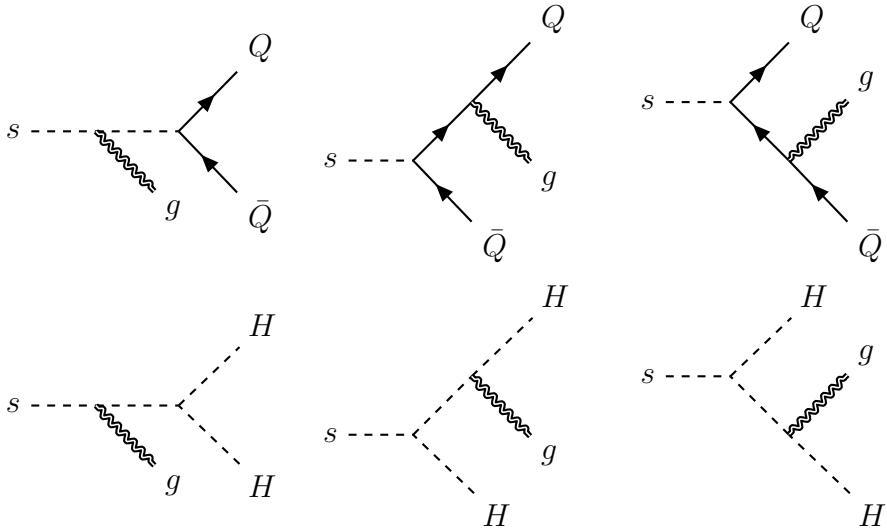
\begin{figure}[t]
	\centering
	\begin{minipage}[c]{.24\textwidth}
		\begin{tikzpicture}
			\begin{feynman}
				\vertex (a) {\(s\)};
				\vertex [right=1.1cm of a] (b) ;
				\vertex [right=1.1cm of b] (c);
				\vertex [above right=1.1cm of c] (d) {\(Q\)};
				\vertex [below right=1.1cm of b] (e) {\(g \)};
				\vertex [below right=1.1cm of c] (f) {\(\bar{Q}\)};
				\diagram* {
					(a) -- [scalar,thick] (b),
					(b) -- [scalar,thick] (c),
					(c) -- [fermion,thick] (d),
					(b) -- [graviton,thick] (e),
					(f) -- [fermion,thick] (c)
				};
				~~~~~~\end{feynman}
		\end{tikzpicture}
\end{minipage}	
\begin{minipage}[c]{.24\textwidth}
		\begin{tikzpicture}
			\begin{feynman}
				\vertex (a) {\(s\)};
				\vertex [right=1.1cm of a] (b) ;
				\vertex [above right=1.1cm of b] (c);
				\vertex [above right=1.1cm of c] (d) {\(Q\)};
				\vertex [below right=1.1cm of c] (e) {\(g \)};
				\vertex [below right=1.1cm of b] (f) {\(\bar{Q}\)};
				\diagram* {
					(a) -- [scalar,thick] (b),
					(b) -- [fermion,thick] (c),
					(c) -- [fermion,thick] (d),
					(c) -- [graviton,thick] (e),
					(f) -- [fermion,thick] (b)
				};
				~~~~~~\end{feynman}
		\end{tikzpicture}
\end{minipage}	
	\begin{minipage}[c]{.24\textwidth}
		\begin{tikzpicture}
			\begin{feynman}
				\vertex (a) {\(s\)};
				\vertex [right=1.1cm of a] (b) ;
				\vertex [above right=1.1cm of b] (c) {\(Q\)};
				\vertex [below right=1.1cm of b] (d) ;
				\vertex [below right=1.1cm of d] (f) {\(\bar{Q}\)};
				\vertex [above right=1.1cm of d] (e) {\(g \)};
				\diagram* {
					(a) -- [scalar,thick] (b),
					(b) -- [fermion,thick] (c),
					(d) -- [fermion,thick] (b),
					(d) -- [graviton,thick] (e),
					(f) -- [fermion,thick] (d)
				};
				~~~~~~\end{feynman}
		\end{tikzpicture}
	\end{minipage}	

	\begin{minipage}[c]{.24\textwidth}
		\begin{tikzpicture}
			\begin{feynman}
				\vertex (a) {\(s\)};
				\vertex [right=1.1cm of a] (b) ;
				\vertex [right=1.1cm of b] (c);
				\vertex [above right=1.1cm of c] (d) {\(H\)};
				\vertex [below right=1.1cm of b] (e) {\(g \)};
				\vertex [below right=1.1cm of c] (f) {\(H\)};
				\diagram* {
					(a) -- [scalar,thick] (b),
					(b) -- [scalar,thick] (c),
					(c) -- [scalar,thick] (d),
					(b) -- [graviton,thick] (e),
					(f) -- [scalar,thick] (c)
				};
				~~~~~~\end{feynman}
		\end{tikzpicture}
\end{minipage}	
\begin{minipage}[c]{.24\textwidth}
		\begin{tikzpicture}
			\begin{feynman}
				\vertex (a) {\(s\)};
				\vertex [right=1.1cm of a] (b) ;
				\vertex [above right=1.1cm of b] (c);
				\vertex [above right=1.1cm of c] (d) {\(H\)};
				\vertex [below right=1.1cm of c] (e) {\(g \)};
				\vertex [below right=1.1cm of b] (f) {\(H\)};
				\diagram* {
					(a) -- [scalar,thick] (b),
					(b) -- [scalar,thick] (c),
					(c) -- [scalar,thick] (d),
					(c) -- [graviton,thick] (e),
					(f) -- [scalar,thick] (b)
				};
				~~~~~~\end{feynman}
		\end{tikzpicture}
\end{minipage}	
	\begin{minipage}[c]{.24\textwidth}
		\begin{tikzpicture}
			\begin{feynman}
				\vertex (a) {\(s\)};
				\vertex [right=1.1cm of a] (b) ;
				\vertex [above right=1.1cm of b] (c) {\(H\)};
				\vertex [below right=1.1cm of b] (d) ;
				\vertex [below right=1.1cm of d] (f) {\(H\)};
				\vertex [above right=1.1cm of d] (e) {\(g \)};
				\diagram* {
					(a) -- [scalar,thick] (b),
					(b) -- [scalar,thick] (c),
					(d) -- [scalar,thick] (b),
					(d) -- [graviton,thick] (e),
					(f) -- [scalar,thick] (d)
				};
				~~~~~~\end{feynman}
		\end{tikzpicture}
	\end{minipage}	
	\caption{Feynman diagrams for graviton bremsstrahlung processes in the decay of the heavy scalar.}
	\label{fig:saxion_decay_diagram}
\end{figure}
The squared amplitude summed over the spin  and polarization of final states is given by \cite{barman2023gravitational}
\begin{align}
|\mathcal{M}_{s \to Q\bar{Q}g}|^2 &= \mathcal{I}_Q \mathcal{C}_Q \frac{\kappa^2 y_Q^2 m_s^2}{8} \frac{ 4 (1 -x_Q)(1 -x_g)(x_Q+x_g-1) - x_g^2 z_Q^2}{x_g^2  (1-x_Q)^2 (1-(x_Q+x_g))^2 } \notag \\
&\quad \times \bigg\{ z_Q^2 \Big[ (x_Q^2 + 3 x_Q x_g + 3 x_g^2/4) - 2(x_Q+x_g) - x_Q x_g(x_Q+x_g) + 1 \Big] \notag \\
&\quad - (1-x_Q)(x_g^2/2 - x_g + 1) \Big[ 1 - (x_Q+x_g) \Big] + x_g^2 z_Q^4/4 \bigg\}.
\end{align}
Here, $\kappa = \sqrt{32 \pi G}$, and we have introduced dimensionless variables $x_Q \equiv 2 E_Q/m_s$, $x_g\equiv 2E_g/m_s$, $z_Q \equiv 2 m_Q / m_s$. The variable $x_g$ lies in the range $ (0 , 1-z_Q^2] $ and $x_Q$ lies in the range $ [1- x_g /2  - \alpha x_g /2 , \; 1- x_g /2  + \alpha x_g /2] $ with $\alpha = \sqrt{1 - z_Q^2 / (1 - x_g)}$.

Taking all the ingredients into the Boltzmann equation Eq. \eqref{Eq.BE}, the GW-spectrum today generated from the graviton bremsstrahlung processes during the decay  $s \to Q+\bar Q +g$  can be expressed as
\begin{equation}
    \begin{aligned}
        \Omega_{GW} h^2 &= h^2 \Omega_\gamma^0 \mathcal{I}_Q \mathcal{C}_Q \frac{ y_Q^2}{16 \pi^2} \frac{m_s^2}{m_{\rm Pl}^2} \int_{t_{m_s}}^{t_D} dt \frac{\rho_s(t)}{\rho_\gamma^0} \frac{a^4(t)}{a^4(t_0)}\frac{m_s}{2} x_g \bigg[  \alpha (1-x_g)\Big[ x_g z_Q^2 + (x_g/2-1)x_g \\
		&- z_Q^4/2 - z_Q^2/2 + 1 \Big]  +z_Q^2 \left[ (5-4 x_g)z_Q^2/4 - (x_g/2-1)^2 - z_Q^4/4 \right] \log \left( \frac{1+\alpha}{1-\alpha} \right) \bigg]  \; ,
    \end{aligned}
\end{equation}
where $m_{\rm Pl} = 1/\sqrt{G}$ is the Planck mass, $\rho_\gamma^0$ is the energy density of photons today, $h^2 \Omega_\gamma^0 = h^2 \rho_\gamma^0/\rho_{cr,0}=2.47 \times 10^{-5}$ \cite{ParticleDataGroup:2024cfk}, $\rho_s = m_s n_s$ is the energy density of $s$ and $a(t)$ is the scale factor of the universe at the time $t$. The graviton energy $E_g$ at time of emission is related to the present-day energy $E_g^0$ via redshift relation: $E_g = E_g^0 a(t_0)/a(t)$.

\begin{figure}[t]
	\centering
	\begin{minipage}
[c]{.45\textwidth}
		\includegraphics[width=\textwidth]{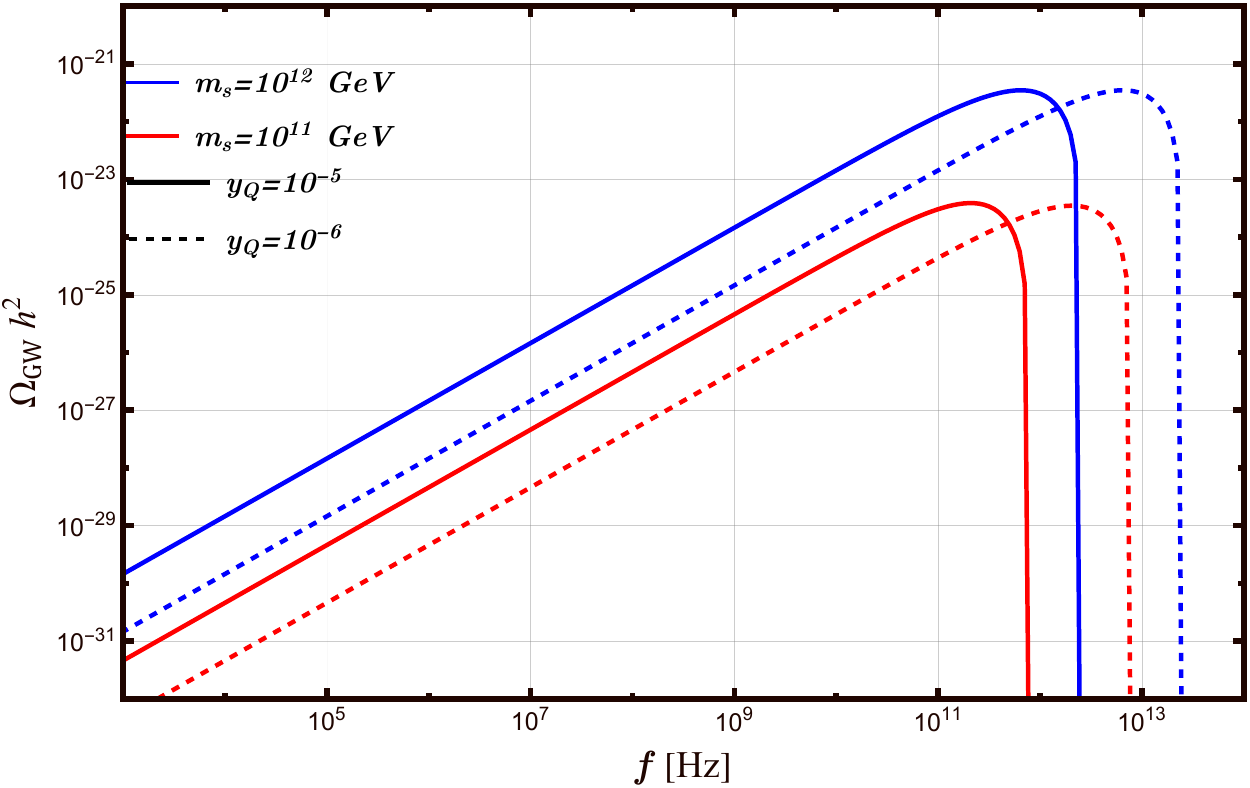}
	\end{minipage}
	\begin{minipage}
		[c]{.45\textwidth}
		\includegraphics[width=\textwidth]{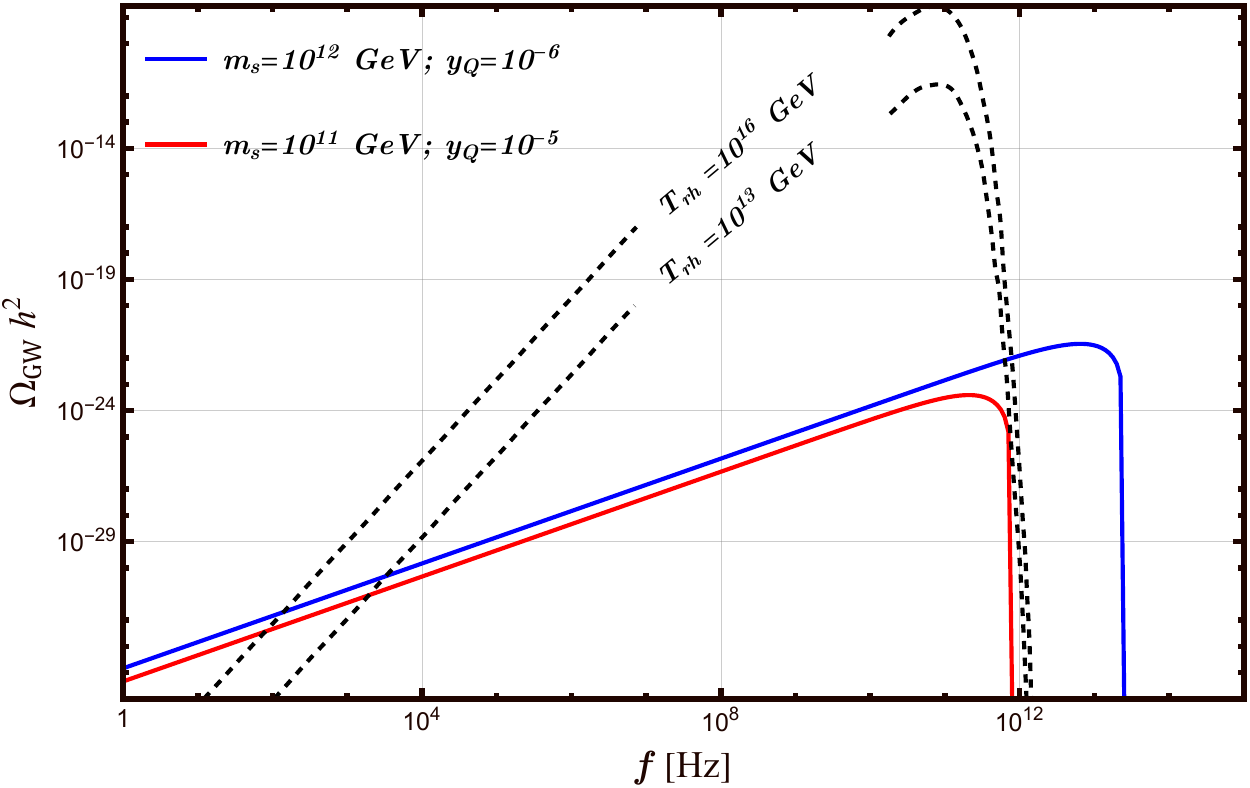}
	\end{minipage}
	\caption{
		GW spectra  from graviton bremsstrahlung in the $s\to Q \bar Q g$ decay channel.
		\textbf{Left Panel:} Dependence of the GW spectrum on  $m_s$ and  $y_Q$. Red and blue curves correspond to $m_s = 10^{12}$ GeV and $10^{11}$ GeV, respectively. Solid and dashed lines represent coupling strengths of $y_Q = 10^{-5}$ and $10^{-6}$, respectively.
		\textbf{Right Panel:} Comparison between the GW spectra from scalar decay (solid lines) and the thermal plasma background (dashed lines) for reheating temperatures $T_{\rm rh} = 10^{16}$ GeV and $T_{\rm rh} = 10^{13}$ GeV. In the $m_s = 10^{12}$ GeV and $y_Q = 10^{-6}$ scenario (red), an early matter-dominated phase is realized, allowing the scalar-induced GW signal to be clearly distinguished from the thermal GW background at high frequencies. Conversely, for $m_s = 10^{11}$ GeV and $y_Q = 10^{-5}$ (blue), no matter-dominated phase occurs;  the high-frequency GW signal is overwhelmed by the thermal plasma contribution, leaving distinguishable spectrum in the low-frequency regime.
		}
	\label{fig:GW_spectrum_saxion_decay}
\end{figure}

The GW-spectra resulting from the graviton bremsstrahlung processes in the decay of saxion into VLQs are shown Fig.~\ref{fig:GW_spectrum_saxion_decay}.
In the left panel , the GW spectra are depicted using different colors for various heavy scalar masses and different line styles for various values of the coupling $y_Q$. To obtain a qualitative insight into the parametric dependence of the GW spectrum on $y_Q$ and $m_s$, we simplify the analysis by adopting the massless limit for $Q$, i.e., $z_Q=0$, and employing an instantaneous decay approximation, wherein the decay is treated as a discrete event at $t = t_D$. Under these approximations, the GW spectrum can be expressed as
\begin{equation}
	\begin{aligned}
	h^2 \Omega_{GW} = &h^2 \Omega_\gamma^0 \mathcal{I}_Q \mathcal{C}_Q \frac{ y_Q^2}{16 \pi^2} \frac{m_s^2}{m_{\rm Pl}^2} \frac{1}{\Gamma_s} \frac{\rho_s(t_D)}{\rho_\gamma^0} \frac{a^4(t_D)}{a^4(t_0)}\frac{m_s}{2} \mathcal{F}(x_g^{D})
	\end{aligned}
\end{equation}
where $\mathcal{F}(x_g^{D}) \equiv x_g^{D} (1-x_g^D)[1-x_g^D + 2(x_g^D)^2]$ with $x_g^D \equiv 2 E_g(t_D)/m_s$. As the universe returns to radiation domination at $t_D$ from the era of early matter domination induced by the heavy scalar, we have $\rho_s(t_D) \simeq \rho_{R}(t_D) \simeq \rho_R(t_{m_s}) a^4(t_{m_s})/a^4(t_D)$, where $\rho_R$ denotes the energy density of radiation. Consequently, the GW spectrum scales with $m_s$ and $y_Q$ as
\begin{equation}
	h^2 \Omega_{GW} \propto m_s^2 \mathcal{F}(x_g^D),
\end{equation}
and is maximized at $x_g^D = 0.65$, which indicates that the peak frequency is 
\begin{equation}
	f_{peak}= 0.65 \times \frac{1}{2\pi} \frac{m_s}{2} \frac{a(t_D)}{a(t_0)} = 2.5 \times 10^{12} \text{Hz} \left(\frac{m_s}{10^{12} \text{GeV}}\right)^{1/2} \left(\frac{y_Q}{10^{-5}}\right)^{-1}.
\end{equation}
These scaling laws for GW-spectrum are in agreement with the figure~\ref{fig:GW_spectrum_saxion_decay}. For a fixed coupling $y_Q$, the increase of mass $m_s$ (illustrated by the transition from blue to red curves) leads to a significant enhancement in the gravitational wave signal. Specifically, the amplitude of the spectrum scales as $m^2_s$, while the peak frequency shifts toward the high frequency regime, following a $\sqrt{m_s}$ dependence. Conversely, when $m_s$ is held fixed (curves of the same color), the amplitude of the spectrum remains unchanged, while the peak frequency shifts according to a $1/y_Q$ scaling law.

In the right panel of the Fig.~\ref{fig:GW_spectrum_saxion_decay}, we compare the GW spectra from $s$ decay with the Cosmic Gravitational Wave Background (CGWB) generated by the thermal plasma \cite{Ringwald:2020ist,Ghiglieri:2015nfa,ghiglieri2020gravitational}. In the scenarios where $s$ triggers an early matter-dominated phase, the GW signals from $s$ decay can be clearly distinguished from the CGWB at high frequency band. Conversely, in the scenarios without a matter-dominated phase, the high-frequency GW signals are overwhelmed by the CGWB, leaving distinguishable spectrum in the low-frequency band, where the GW signals are beyond the reach of current GW observatories.
We have ignored the effects of an early matter-dominated phase on the CGWB spectrum in this comparison. A detailed analysis of these effects can be found in \cite{murayama2025observing}.

\subsubsection{Higgs decay channel}
We now turn to the scenario $s\to H+H +g$, which is similar to the graviton bremsstrahlung process during the decay of the Higgs triplet into a pair of SM Higgs doublets within the Type-II Seesaw model\cite{Wang:2025mtq}. The relevant Feynman diagrams for the graviton bremsstrahlung processes  are shown in the second row of the Fig.~\ref{fig:saxion_decay_diagram}.

The squared amplitude summed over the polarizations is given by
\begin{equation}
	|\mathcal{M}_{s \to HHg}|^2 = \frac{\kappa^2}{2} \mu_{\Phi H}^2 \frac{\left(1- x_g\right)^2}{x_g^2},
\end{equation}
where $x_g$ is in the range $(0,1]$. Consequently, the GW-spectrum today from the graviton bremsstrahlung processes in the decay of saxion into Higgs doublets can be expressed as
\begin{equation}
	h^2 \Omega_{GW} = h^2 \Omega_\gamma^0 \frac{2}{16 \pi^2} \frac{\mu_{\Phi H}^2}{m_{\rm Pl}^2} \int_{t_{PQ}}^{t_D} dt \frac{\rho_s(t)}{\rho_\gamma^0} \frac{a^4(t)}{a^4(t_0)}\frac{m_s}{2} x_g (1 - x_g)^2,
\end{equation}
where a factor of 2 comes from the fact that final states are $SU(2)$ doublets.

Under the instantaneous decay approximation, the GW spectrum scales as
\begin{equation}
	h^2 \Omega_{GW} \sim m_s^2 x_g^D (1 -x_g^D)^2.
\end{equation}
Consequently, the peak frequency is expected to follow $f_{\rm peak} \sim m_s^{3/2}/\mu_{\Phi H}$.  It should be mentioned that the interaction $\lambda_{H\Phi} (\Phi^\dagger \Phi )(H^\dagger H)$ contributes to a mass term of the SM Higgs, that will forbid the electroweak symmetry breaking (EWSB) or trigger an earlier EWSB for large $|\lambda_{H\Phi}|$, which is conflict with the traditional EWSB scenario. On the other hand a tiny $\lambda_{H\Phi}$ will suppress the induced GW spectrum. Thus we can safely neglect the GW induced by this process.

\subsection{GW-spectrum from the vector-like quark decay}
\begin{figure}[t]
	\centering
	\begin{minipage}[c]{.24\textwidth}
		\begin{tikzpicture}
			\begin{feynman}
				\vertex (a) {\(Q\)};
				\vertex [right=1.1cm of a] (b) ;
				\vertex [right=1.1cm of b] (c);
				\vertex [above right=1.1cm of c] (d) {\(q\)};
				\vertex [below right=1.1cm of b] (e) {\(g \)};
				\vertex [below right=1.1cm of c] (f) {\(H\)};
				\diagram* {
					(a) -- [fermion,thick] (b),
					(b) -- [fermion,thick] (c),
					(c) -- [fermion,thick] (d),
					(b) -- [graviton,thick] (e),
					(f) -- [scalar,thick] (c)
				};
				~~~~~~\end{feynman}
		\end{tikzpicture}
\end{minipage}	
    \begin{minipage}[c]{.24\textwidth}
		\begin{tikzpicture}
			\begin{feynman}
				\vertex (a) {\(Q\)};
				\vertex [right=1.1cm of a] (b) ;
				\vertex [above right=1.1cm of b] (c);
				\vertex [above right=1.1cm of c] (d) {\(q\)};
				\vertex [below right=1.1cm of c] (e) {\(g \)};
				\vertex [below right=1.1cm of b] (f) {\(H\)};
				\diagram* {
					(a) -- [fermion,thick] (b),
					(b) -- [fermion,thick] (c),
					(c) -- [fermion,thick] (d),
					(c) -- [graviton,thick] (e),
					(b) -- [scalar,thick] (f)
				};
				~~~~~~\end{feynman}
		\end{tikzpicture}
\end{minipage}	
	\begin{minipage}[c]{.24\textwidth}
		\begin{tikzpicture}
			\begin{feynman}
				\vertex (a) {\(Q\)};
				\vertex [right=1.1cm of a] (b) ;
				\vertex [above right=1.1cm of b] (c) {\(q\)};
				\vertex [below right=1.1cm of b] (d) ;
				\vertex [below right=1.1cm of d] (f) {\(H\)};
				\vertex [above right=1.1cm of d] (e) {\(g \)};
				\diagram* {
					(a) -- [fermion,thick] (b),
					(b) -- [fermion,thick] (c),
					(b) -- [scalar,thick] (d),
					(d) -- [graviton,thick] (e),
					(f) -- [scalar,thick] (d)
				};
				~~~~~~\end{feynman}
		\end{tikzpicture}
	\end{minipage}	
	\caption{Feynman diagrams for graviton bremsstrahlung in the decay of the vector-like quark. }
	\label{fig:feynman}
\end{figure}
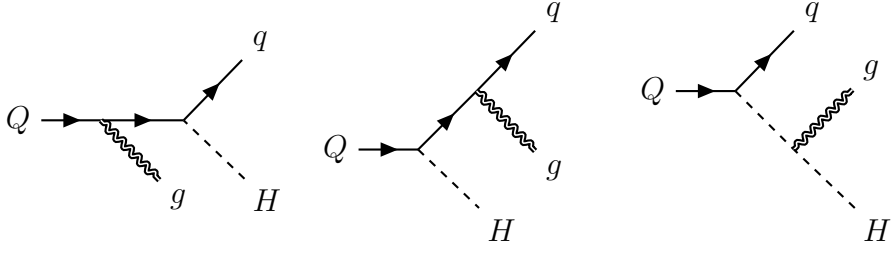

In this section, we calculate the graviton bremsstrahlung from the decay  $Q \to q H g$ with $g$. The corresponding Feynman diagrams are shown in Fig.~\ref{fig:feynman}.

The squared amplitude summed over the spin and polarization of final states is given by
\begin{equation}
    |\mathcal{M}_{Q \to qHg}|^2 =y_{Qq}^2 \kappa ^2 \frac{ m_Q^2 (x_g-1) (x_g x_q+x_g-2)}{8 x_g^2}
\end{equation}
where $\kappa = \sqrt{32 \pi G}$, and $x_i = 2 E_i / m_Q$ with $i=g,q$ are the dimensionless energy variables of the graviton and the SM quark, respectively. The variable $x_g$ is in the range $(0,1]$ and $x_q$ is in the range $[1 - x_g, 1]$.

\begin{figure}[t]
	\centering
	\includegraphics[width=0.75\textwidth]{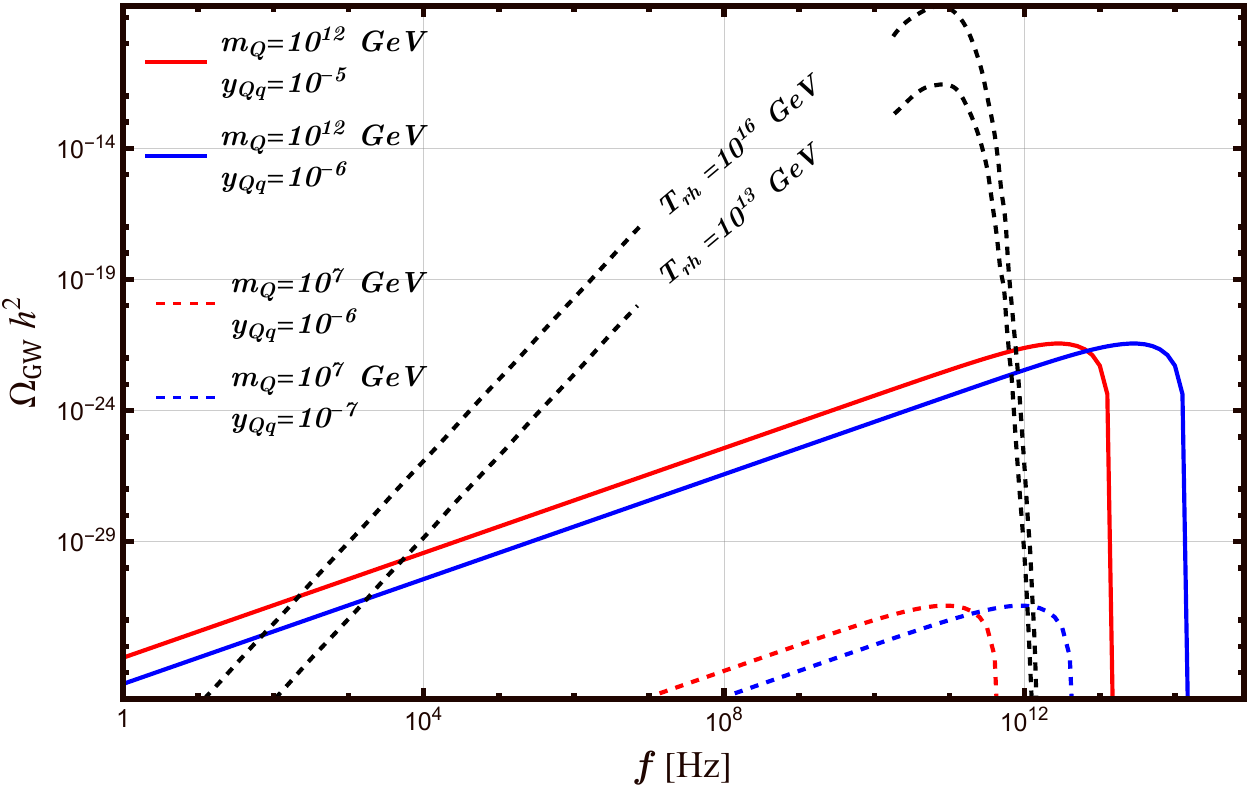}
	\caption{Gravitational wave spectra arising from graviton bremsstrahlung in VLQ decays. \textbf{Black dashed line} is the CGWB for reheating temperatures $T_{\rm rh} = 10^{16}$ GeV and $T_{\rm rh} = 10^{13}$ GeV. \textbf{Colored dashed lines} (red: $m_Q=10^7$ GeV, $y_{Qq}=10^{-6}$; blue: $m_Q=10^7$ GeV, $y_{Qq}=10^{-7}$) represent scenarios where $s$ predominantly decay into VLQs. In these cases, the small coupling $y_Q$ required to distinguish the signal from the CGWB necessitates relatively small VLQ masses, leading to a suppressed GW spectrum. \textbf{Solid lines} (red: $m_Q=10^{12}$ GeV, $y_{Qq}=10^{-5}$; blue: $m_Q=10^{12}$ GeV, $y_{Qq}=10^{-6}$) correspond to scenarios where $s\to Q\bar Q$ is forbidden, which permit larger VLQ masses. These high-frequency signals are clearly distinguishable from the CGWB, provided that the VLQs trigger an early matter-dominated era.}
	\label{fig:GW_spectrum_heavyquark_decay}
\end{figure}

Therefore, the GW energy spectrum today from the VLQ decay can be expressed as
\begin{equation}
    \Omega_{GW} h^2 = h^2 \Omega_\gamma^0  \frac{m_Q^2}{m_{\rm Pl}^2} \frac{y_{Qq}^2}{128 \pi^2} \int_{t_{m_Q}}^{t_D} dt \frac{\rho_Q(t)}{\rho_\gamma^0} \frac{a^4(t)}{a^4(t_0)} E_g (1-2E_g/m_Q)(2- 2E_g/m_Q)^2,
\end{equation}
where $\rho_Q = m_Q n_Q$ is the energy density of the VLQ. 
In the scenario where heavy scalar primarily decays into VLQs, distinguishing the resulting GW signal from the CGWB requires a small coupling $y_Q$. This requirement implies a VLQ mass significantly smaller than that of the heavy scalar. 
Since the GW amplitude scales as $m_Q^2$, the signal from the VLQ decay is notably suppressed, as indicated by the dashed red and blue lines in Fig.~\ref{fig:GW_spectrum_heavyquark_decay}. Conversely,  larger value of $m_Q$ are permitted if $s \to Q\bar Q$ is kinetically forbidden. In this case, the GW signal from the VLQ decay reaches a magnitude comparable to the heavy scalar induced  GW signal. Furthermore, when VLQ dynamics trigger an early matter-dominated universe, the GW signals become distinguishable from the CGWB, as illustrated by the solid blue and red lines in Fig.~\ref{fig:GW_spectrum_heavyquark_decay}.

Under the instantaneous decay approximation, the parametric dependence of this GW spectrum is given by
   \begin{equation}
	 h^2 \Omega_{GW} \propto m_Q^2 \mathcal{F}(x_g^D) , \label{Eq.prop}
   \end{equation}
where $\mathcal{F}(x_g^D) \equiv x_g^D (1-x_g^D)(2-x_g^D)^2$ with $x_g^D\equiv 2E_g/m_Q$. The peak frequency thus is given by
\begin{equation}
	f_{peak} = 0.36 \times \frac{1}{2\pi} \frac{m_Q}{2} \frac{a(t_D)}{a(t_0)} = 4.9 \times 10^{12} \text{Hz} \left(\frac{m_Q}{10^{12} \text{GeV}}\right)^{1/2} \left(\frac{y_{Qq}}{10^{-5}}\right)^{-1}.
\end{equation}
Hence, the dependence of the GW spectrum on variations of $m_Q$ and $y_{Qq}$ is analogous to that observed in the heavy scalar decay scenario.

\section{Conclusion}


The direct detection of the QCD axion in the laboratory and the subsequent discrimination between its possible UV completions, such as the KSVZ and DFSZ models, represent paramount challenges in fundamental physics. A critical obstacle is that even upon the successful discovery of the axion via traditional electromagnetic channels, its microphysical origin may remain obscured. This ambiguity arises because key observables, like the axion's mass and photon coupling, can be similar across different models, and because cosmic backgrounds—the electromagnetic ``fog" of the universe—limit the clarity of high-energy electromagnetic signatures from early-universe processes.
To bypass this impasse, we propose ultra-high-frequency gravitational waves (UHF-GWs) as a novel and complementary cosmological probe. GWs, being notoriously weak interacting field, travel unimpeded across the cosmos, carrying pristine information from the earliest and most energetic epochs, inaccessible to electromagnetic astronomy.
In this work, we have performed a dedicated calculation of the UV physics induced GW spectrum within the framework of the KSVZ axion model. We investigate a potent source of UHF-GWs: graviton bremsstrahlung emitted during the decays of the scalar partner of the axion and the heavy VLQ intrinsic to the KSVZ construction. Our analysis reveals that this mechanism can generate a distinctive, high-frequency stochastic GW background, with a spectrum typically peaking in the ultra-high frequency range.
Crucially, the predicted spectral shape, peak frequency, and amplitude are directly tied to the unique particle content and couplings of the KSVZ model—namely, the masses and Yukawa couplings of the VLQ and the heavy scalar. When contrasted with the GW spectrum predicted by alternative frameworks like the DFSZ model, we identify model-discriminating features. These stark differences in the UHF-GW signature arise from the fundamentally distinct UV physics of each model, offering a clear observational fingerprint.
Therefore, the future detection of the UHF-GW background could provide an indirect yet powerful diagnostic tool for axion model selection. While the direct detection of UHF-GWs presents significant technological challenges, next-generation experiments utilizing advanced resonant cavities, quantum sensors, and cryogenic techniques are rapidly evolving. Our work provides a concrete theoretical target for these endeavors, charting a path toward using multi-messenger astronomy—combining GW data with direct axion searches and astrophysical bounds—to finally lift the veil on the axion's ultraviolet origin.

\section*{Acknowledgements}

This work was supported by the National Natural Science Foundation of China (NSFC) (Grants No. 12447105, No. 11775025 and No. 12175027), and the Fundamental Research Funds for the Central Universities.

\bibliographystyle{elsarticle-num}
\bibliography{references}

\begin{thebibliography}{10}
\expandafter\ifx\csname url\endcsname\relax
  \def\url#1{\texttt{#1}}\fi
\expandafter\ifx\csname urlprefix\endcsname\relax\def\urlprefix{URL }\fi
\expandafter\ifx\csname href\endcsname\relax
  \def\href#1#2{#2} \def\path#1{#1}\fi

\bibitem{Peccei:1977hh}
R.~D. Peccei, H.~R. Quinn, {CP Conservation in the Presence of Instantons},
  Phys. Rev. Lett. 38 (1977) 1440--1443.
\newblock \href {https://doi.org/10.1103/PhysRevLett.38.1440}
  {\path{doi:10.1103/PhysRevLett.38.1440}}.

\bibitem{Peccei:1977ur}
R.~D. Peccei, H.~R. Quinn, {Constraints Imposed by CP Conservation in the
  Presence of Instantons}, Phys. Rev. D 16 (1977) 1791--1797.
\newblock \href {https://doi.org/10.1103/PhysRevD.16.1791}
  {\path{doi:10.1103/PhysRevD.16.1791}}.

\bibitem{Weinberg:1977ma}
S.~Weinberg, {A New Light Boson?}, Phys. Rev. Lett. 40 (1978) 223--226.
\newblock \href {https://doi.org/10.1103/PhysRevLett.40.223}
  {\path{doi:10.1103/PhysRevLett.40.223}}.

\bibitem{Wilczek:1977pj}
F.~Wilczek, {Problem of Strong $P$ and $T$ Invariance in the Presence of
  Instantons}, Phys. Rev. Lett. 40 (1978) 279--282.
\newblock \href {https://doi.org/10.1103/PhysRevLett.40.279}
  {\path{doi:10.1103/PhysRevLett.40.279}}.

\bibitem{Vafa:1984xg}
C.~Vafa, E.~Witten, {Parity Conservation in QCD}, Phys. Rev. Lett. 53 (1984)
  535.
\newblock \href {https://doi.org/10.1103/PhysRevLett.53.535}
  {\path{doi:10.1103/PhysRevLett.53.535}}.

\bibitem{Kim:1979if}
J.~E. Kim, {Weak Interaction Singlet and Strong CP Invariance}, Phys. Rev.
  Lett. 43 (1979) 103.
\newblock \href {https://doi.org/10.1103/PhysRevLett.43.103}
  {\path{doi:10.1103/PhysRevLett.43.103}}.

\bibitem{Shifman:1979if}
M.~A. Shifman, A.~I. Vainshtein, V.~I. Zakharov, {Can Confinement Ensure
  Natural CP Invariance of Strong Interactions?}, Nucl. Phys. B 166 (1980)
  493--506.
\newblock \href {https://doi.org/10.1016/0550-3213(80)90209-6}
  {\path{doi:10.1016/0550-3213(80)90209-6}}.

\bibitem{Dine:1981rt}
M.~Dine, W.~Fischler, M.~Srednicki, {A Simple Solution to the Strong CP Problem
  with a Harmless Axion}, Phys. Lett. B 104 (1981) 199--202.
\newblock \href {https://doi.org/10.1016/0370-2693(81)90590-6}
  {\path{doi:10.1016/0370-2693(81)90590-6}}.

\bibitem{Zhitnitsky:1980tq}
A.~R. Zhitnitsky, {On Possible Suppression of the Axion Hadron Interactions.
  (In Russian)}, Sov. J. Nucl. Phys. 31 (1980) 260.

\bibitem{Kim:1986ax}
J.~E. Kim, {Light Pseudoscalars, Particle Physics and Cosmology}, Phys. Rept.
  150 (1987) 1--177.
\newblock \href {https://doi.org/10.1016/0370-1573(87)90017-2}
  {\path{doi:10.1016/0370-1573(87)90017-2}}.

\bibitem{LIGOScientific:2016aoc}
B.~P. Abbott, et~al., {Observation of Gravitational Waves from a Binary Black
  Hole Merger}, Phys. Rev. Lett. 116~(6) (2016) 061102.
\newblock \href {http://arxiv.org/abs/1602.03837} {\path{arXiv:1602.03837}},
  \href {https://doi.org/10.1103/PhysRevLett.116.061102}
  {\path{doi:10.1103/PhysRevLett.116.061102}}.

\bibitem{Ghiglieri:2015nfa}
J.~Ghiglieri, M.~Laine, {Gravitational wave background from Standard Model
  physics: Qualitative features}, JCAP 07 (2015) 022.
\newblock \href {http://arxiv.org/abs/1504.02569} {\path{arXiv:1504.02569}},
  \href {https://doi.org/10.1088/1475-7516/2015/07/022}
  {\path{doi:10.1088/1475-7516/2015/07/022}}.

\bibitem{Ringwald:2022xif}
A.~Ringwald, C.~Tamarit, {Revealing the cosmic history with gravitational
  waves}, Phys. Rev. D 106~(6) (2022) 063027.
\newblock \href {http://arxiv.org/abs/2203.00621} {\path{arXiv:2203.00621}},
  \href {https://doi.org/10.1103/PhysRevD.106.063027}
  {\path{doi:10.1103/PhysRevD.106.063027}}.

\bibitem{Ghiglieri:2022rfp}
J.~Ghiglieri, J.~Sch{\"u}tte-Engel, E.~Speranza, {Freezing-in gravitational
  waves}, Phys. Rev. D 109~(2) (2024) 023538.
\newblock \href {http://arxiv.org/abs/2211.16513} {\path{arXiv:2211.16513}},
  \href {https://doi.org/10.1103/PhysRevD.109.023538}
  {\path{doi:10.1103/PhysRevD.109.023538}}.

\bibitem{Ghiglieri:2024ghm}
J.~Ghiglieri, M.~Laine, J.~Sch{\"u}tte-Engel, E.~Speranza, {Double-graviton
  production from Standard Model plasma}, JCAP 04 (2024) 062.
\newblock \href {http://arxiv.org/abs/2401.08766} {\path{arXiv:2401.08766}},
  \href {https://doi.org/10.1088/1475-7516/2024/04/062}
  {\path{doi:10.1088/1475-7516/2024/04/062}}.

\bibitem{Ringwald:2020ist}
A.~Ringwald, J.~Sch{\"u}tte-Engel, C.~Tamarit, {Gravitational Waves as a Big
  Bang Thermometer}, JCAP 03 (2021) 054.
\newblock \href {http://arxiv.org/abs/2011.04731} {\path{arXiv:2011.04731}},
  \href {https://doi.org/10.1088/1475-7516/2021/03/054}
  {\path{doi:10.1088/1475-7516/2021/03/054}}.

\bibitem{barman2023gravitational}
B.~Barman, N.~Bernal, Y.~Xu, {\'O}.~Zapata, Gravitational wave from graviton
  bremsstrahlung during reheating, Journal of Cosmology and Astroparticle
  Physics 2023~(05) (2023) 019.

\bibitem{Xu:2025wjq}
X.-J. Xu, Y.~Xu, Q.~Yin, J.~Zhu, {Full-Spectrum Analysis of Gravitational Wave
  Production from Inflation to Reheating} (5 2025).
\newblock \href {http://arxiv.org/abs/2505.08868} {\path{arXiv:2505.08868}}.

\bibitem{huang2019stochastic}
D.~Huang, L.~Yin, Stochastic gravitational waves from inflaton decays, Physical
  Review D 100~(4) (2019) 043538.

\bibitem{Barman:2023rpg}
B.~Barman, N.~Bernal, Y.~Xu, {\'O}.~Zapata, {Bremsstrahlung-induced
  gravitational waves in monomial potentials during reheating}, Phys. Rev. D
  108~(8) (2023) 083524.
\newblock \href {http://arxiv.org/abs/2305.16388} {\path{arXiv:2305.16388}},
  \href {https://doi.org/10.1103/PhysRevD.108.083524}
  {\path{doi:10.1103/PhysRevD.108.083524}}.

\bibitem{Bernal:2023wus}
N.~Bernal, S.~Cl{\'e}ry, Y.~Mambrini, Y.~Xu, {Probing reheating with graviton
  bremsstrahlung}, JCAP 01 (2024) 065.
\newblock \href {http://arxiv.org/abs/2311.12694} {\path{arXiv:2311.12694}},
  \href {https://doi.org/10.1088/1475-7516/2024/01/065}
  {\path{doi:10.1088/1475-7516/2024/01/065}}.

\bibitem{Xu:2024fjl}
Y.~Xu, {Ultra-high frequency gravitational waves from scattering,
  Bremsstrahlung and decay during reheating}, JHEP 10 (2024) 174.
\newblock \href {http://arxiv.org/abs/2407.03256} {\path{arXiv:2407.03256}},
  \href {https://doi.org/10.1007/JHEP10(2024)174}
  {\path{doi:10.1007/JHEP10(2024)174}}.

\bibitem{Xu:2024xmw}
Y.~Xu, {Gravitational Wave from Graviton Bremsstrahlung during Reheating}, PoS
  CORFU2023 (2024) 047.
\newblock \href {https://doi.org/10.22323/1.463.0047}
  {\path{doi:10.22323/1.463.0047}}.

\bibitem{Bernal:2024jim}
N.~Bernal, Y.~Xu, {Thermal gravitational waves during reheating}, JHEP 01
  (2025) 137.
\newblock \href {http://arxiv.org/abs/2410.21385} {\path{arXiv:2410.21385}},
  \href {https://doi.org/10.1007/JHEP01(2025)137}
  {\path{doi:10.1007/JHEP01(2025)137}}.

\bibitem{Bernal:2025lxp}
N.~Bernal, Q.-f. Wu, X.-J. Xu, Y.~Xu, {Pre-thermalized Gravitational Waves} (3
  2025).
\newblock \href {http://arxiv.org/abs/2503.10756} {\path{arXiv:2503.10756}}.

\bibitem{Tokareva:2023mrt}
A.~Tokareva, {Gravitational waves from inflaton decay and bremsstrahlung},
  Phys. Lett. B 853 (2024) 138695.
\newblock \href {http://arxiv.org/abs/2312.16691} {\path{arXiv:2312.16691}},
  \href {https://doi.org/10.1016/j.physletb.2024.138695}
  {\path{doi:10.1016/j.physletb.2024.138695}}.

\bibitem{Kanemura:2023pnv}
S.~Kanemura, K.~Kaneta, {Gravitational waves from particle decays during
  reheating}, Phys. Lett. B 855 (2024) 138807.
\newblock \href {http://arxiv.org/abs/2310.12023} {\path{arXiv:2310.12023}},
  \href {https://doi.org/10.1016/j.physletb.2024.138807}
  {\path{doi:10.1016/j.physletb.2024.138807}}.

\bibitem{Montefalcone:2025gxx}
G.~Montefalcone, B.~Shams Es~Haghi, T.~Xu, K.~Freese, {Thermal Gravitons from
  Warm Inflation} (7 2025).
\newblock \href {http://arxiv.org/abs/2507.08739} {\path{arXiv:2507.08739}}.

\bibitem{Ema:2020ggo}
Y.~Ema, R.~Jinno, K.~Nakayama, {High-frequency Graviton from Inflaton
  Oscillation}, JCAP 09 (2020) 015.
\newblock \href {http://arxiv.org/abs/2006.09972} {\path{arXiv:2006.09972}},
  \href {https://doi.org/10.1088/1475-7516/2020/09/015}
  {\path{doi:10.1088/1475-7516/2020/09/015}}.

\bibitem{Klose:2022knn}
P.~Klose, M.~Laine, S.~Procacci, {Gravitational wave background from
  non-Abelian reheating after axion-like inflation}, JCAP 05 (2022) 021.
\newblock \href {http://arxiv.org/abs/2201.02317} {\path{arXiv:2201.02317}},
  \href {https://doi.org/10.1088/1475-7516/2022/05/021}
  {\path{doi:10.1088/1475-7516/2022/05/021}}.

\bibitem{Klose:2022rxh}
P.~Klose, M.~Laine, S.~Procacci, {Gravitational wave background from vacuum and
  thermal fluctuations during axion-like inflation}, JCAP 12 (2022) 020.
\newblock \href {http://arxiv.org/abs/2210.11710} {\path{arXiv:2210.11710}},
  \href {https://doi.org/10.1088/1475-7516/2022/12/020}
  {\path{doi:10.1088/1475-7516/2022/12/020}}.

\bibitem{An:2022cce}
H.~An, K.-F. Lyu, L.-T. Wang, S.~Zhou, {Gravitational waves from an inflation
  triggered first-order phase transition}, JHEP 06 (2022) 050.
\newblock \href {http://arxiv.org/abs/2201.05171} {\path{arXiv:2201.05171}},
  \href {https://doi.org/10.1007/JHEP06(2022)050}
  {\path{doi:10.1007/JHEP06(2022)050}}.

\bibitem{Hu:2025xdt}
X.-H. Hu, Y.-L. Zhou, {Gravitational waves of GUT phase transition during
  inflation}, Phys. Rev. D 111~(11) (2025) 115003.
\newblock \href {http://arxiv.org/abs/2501.01491} {\path{arXiv:2501.01491}},
  \href {https://doi.org/10.1103/kbzd-kgdr} {\path{doi:10.1103/kbzd-kgdr}}.

\bibitem{Chao:2017ilw}
W.~Chao, W.-F. Cui, H.-K. Guo, J.~Shu, {Gravitational wave imprint of new
  symmetry breaking}, Chin. Phys. C 44~(12) (2020) 123102.
\newblock \href {http://arxiv.org/abs/1707.09759} {\path{arXiv:1707.09759}},
  \href {https://doi.org/10.1088/1674-1137/abb4cb}
  {\path{doi:10.1088/1674-1137/abb4cb}}.

\bibitem{Chao:2017vrq}
W.~Chao, H.-K. Guo, J.~Shu, {Gravitational Wave Signals of Electroweak Phase
  Transition Triggered by Dark Matter}, JCAP 09 (2017) 009.
\newblock \href {http://arxiv.org/abs/1702.02698} {\path{arXiv:1702.02698}},
  \href {https://doi.org/10.1088/1475-7516/2017/09/009}
  {\path{doi:10.1088/1475-7516/2017/09/009}}.

\bibitem{Chao:2023lox}
W.~Chao, J.-j. Feng, H.-k. Guo, T.~Li, {Oscillations of ultralight dark photon
  into gravitational waves}, Nucl. Phys. B 1009 (2024) 116740.
\newblock \href {http://arxiv.org/abs/2312.04017} {\path{arXiv:2312.04017}},
  \href {https://doi.org/10.1016/j.nuclphysb.2024.116740}
  {\path{doi:10.1016/j.nuclphysb.2024.116740}}.

\bibitem{Wang:2025lmf}
Y.~Wang, W.~Chao, {Gravitational Wave Spectrum from the Production of Dark
  Matter via the freeze-in Mechanism} (8 2025).
\newblock \href {http://arxiv.org/abs/2508.10665} {\path{arXiv:2508.10665}}.

\bibitem{Konar:2025iuk}
P.~Konar, S.~Show, {Unraveling Freeze-in Dark matter through the echoes of
  gravitational waves} (6 2025).
\newblock \href {http://arxiv.org/abs/2506.08106} {\path{arXiv:2506.08106}}.

\bibitem{Wang:2025mtq}
Y.~Wang, W.~Chao, {Testing the type-II seesaw mechanism with gravitational
  waves} (10 2025).
\newblock \href {http://arxiv.org/abs/2510.26235} {\path{arXiv:2510.26235}}.

\bibitem{Caprini:2009yp}
C.~Caprini, R.~Durrer, G.~Servant, {The stochastic gravitational wave
  background from turbulence and magnetic fields generated by a first-order
  phase transition}, JCAP 12 (2009) 024.
\newblock \href {http://arxiv.org/abs/0909.0622} {\path{arXiv:0909.0622}},
  \href {https://doi.org/10.1088/1475-7516/2009/12/024}
  {\path{doi:10.1088/1475-7516/2009/12/024}}.

\bibitem{Hindmarsh:2017gnf}
M.~Hindmarsh, S.~J. Huber, K.~Rummukainen, D.~J. Weir, {Shape of the acoustic
  gravitational wave power spectrum from a first order phase transition}, Phys.
  Rev. D 96~(10) (2017) 103520, [Erratum: Phys.Rev.D 101, 089902 (2020)].
\newblock \href {http://arxiv.org/abs/1704.05871} {\path{arXiv:1704.05871}},
  \href {https://doi.org/10.1103/PhysRevD.96.103520}
  {\path{doi:10.1103/PhysRevD.96.103520}}.

\bibitem{Athron:2023xlk}
P.~Athron, C.~Bal{\'a}zs, A.~Fowlie, L.~Morris, L.~Wu, {Cosmological phase
  transitions: From perturbative particle physics to gravitational waves},
  Prog. Part. Nucl. Phys. 135 (2024) 104094.
\newblock \href {http://arxiv.org/abs/2305.02357} {\path{arXiv:2305.02357}},
  \href {https://doi.org/10.1016/j.ppnp.2023.104094}
  {\path{doi:10.1016/j.ppnp.2023.104094}}.

\bibitem{Chao:2021xqv}
W.~Chao, H.-K. Guo, X.-F. Li, {First order color symmetry breaking and
  restoration triggered by electroweak symmetry non-restoration}, Phys. Lett. B
  849 (2024) 138430.
\newblock \href {http://arxiv.org/abs/2112.13580} {\path{arXiv:2112.13580}},
  \href {https://doi.org/10.1016/j.physletb.2023.138430}
  {\path{doi:10.1016/j.physletb.2023.138430}}.

\bibitem{PhysRevLett.43.103}
J.~E. Kim, Weak-interaction singlet and strong $\mathrm{CP}$ invariance, Phys.
  Rev. Lett. 43 (1979) 103--107.
\newblock \href {https://doi.org/10.1103/PhysRevLett.43.103}
  {\path{doi:10.1103/PhysRevLett.43.103}}.

\bibitem{SHIFMAN1980493}
M.~Shifman, A.~Vainshtein, V.~Zakharov, Can confinement ensure natural cp
  invariance of strong interactions?, Nuclear Physics B 166~(3) (1980)
  493--506.
\newblock \href {https://doi.org/https://doi.org/10.1016/0550-3213(80)90209-6}
  {\path{doi:https://doi.org/10.1016/0550-3213(80)90209-6}}.

\bibitem{DiLuzio:2017pfr}
L.~Di~Luzio, F.~Mescia, E.~Nardi, {Window for preferred axion models}, Phys.
  Rev. D 96~(7) (2017) 075003.
\newblock \href {http://arxiv.org/abs/1705.05370} {\path{arXiv:1705.05370}},
  \href {https://doi.org/10.1103/PhysRevD.96.075003}
  {\path{doi:10.1103/PhysRevD.96.075003}}.

\bibitem{Ringwald:2015dsf}
A.~Ringwald, K.~Saikawa, {Axion dark matter in the post-inflationary
  Peccei-Quinn symmetry breaking scenario}, Phys. Rev. D 93~(8) (2016) 085031,
  [Addendum: Phys.Rev.D 94, 049908 (2016)].
\newblock \href {http://arxiv.org/abs/1512.06436} {\path{arXiv:1512.06436}},
  \href {https://doi.org/10.1103/PhysRevD.93.085031}
  {\path{doi:10.1103/PhysRevD.93.085031}}.

\bibitem{Baumann_2022}
D.~Baumann, Cosmology, Cambridge University Press, 2022.

\bibitem{Hooper:2024avz}
D.~Hooper, {Particle Cosmology and Astrophysics}, Princeton University Press,
  2024.

\bibitem{Kolb:1990vq}
E.~W. Kolb, M.~S. Turner, {The Early Universe}, Vol.~69, Taylor and Francis,
  2019.
\newblock \href {https://doi.org/10.1201/9780429492860}
  {\path{doi:10.1201/9780429492860}}.

\bibitem{murayama2025observing}
H.~Murayama, B.~Noether, J.~Sch{\"u}tte-Engel, Observing leptogenesis in action
  with gravitational waves, arXiv preprint arXiv:2506.15772 (2025).

\bibitem{Maggiore_book}
M.~Maggiore, \href{https://doi.org/10.1093/oso/9780198520733.001.0001}{A Modern
  Introduction to Quantum Field Theory}, Oxford University Press, 2004.
\newblock \href {https://doi.org/10.1093/oso/9780198520733.001.0001}
  {\path{doi:10.1093/oso/9780198520733.001.0001}}.
\newline\urlprefix\url{https://doi.org/10.1093/oso/9780198520733.001.0001}

\bibitem{ParticleDataGroup:2024cfk}
S.~Navas, et~al., {Review of particle physics}, Phys. Rev. D 110~(3) (2024)
  030001.
\newblock \href {https://doi.org/10.1103/PhysRevD.110.030001}
  {\path{doi:10.1103/PhysRevD.110.030001}}.

\bibitem{ghiglieri2020gravitational}
J.~Ghiglieri, G.~Jackson, M.~Laine, Y.~Zhu, Gravitational wave background from
  standard model physics: Complete leading order, Journal of High Energy
  Physics 2020~(7) (2020) 1--30.

\end{thebibliography}
\end{document}